\documentclass[amssymb,amsmath,pra,twocolumn,floatfix,showpacs]{revtex4-1}

\usepackage[utf8x]{inputenc}
\usepackage{graphicx}
\usepackage[matrix,frame,arrow]{xy}
\usepackage{amsmath}

\makeatletter

\begin{document}

\title{Fault-tolerant quantum error correction code conversion}

\author{Charles D. Hill} 

\author{Austin G. Fowler}

\author{David S. Wang}

\author{Lloyd C.~L.~Hollenberg}
 
\affiliation{Centre for Quantum
  Computation and Communication Technology, School of Physics,
  University of Melbourne, Melbourne, Victoria 3010, Australia.}

\begin{abstract}
In this paper we demonstrate how data encoded in a five-qubit quantum
error correction code can be converted, fault-tolerantly, into a seven-qubit Steane code. This is achieved by progressing through a series of
codes, each of which fault-tolerantly corrects at least one
error. Throughout the conversion the encoded qubit remains
protected. We found, through computational search, that the method
used to convert between codes given in this paper is optimal.
\end{abstract}


\def\ket#1{\mid\!#1\,\rangle}

\maketitle

\section{Introduction}

In order for a quantum computer to be able to achieve large scale
quantum computing, quantum error correction codes (QECC) will be
required to mitigate the effects of errors due to decoherence and/or
control imprecision. Without quantum error correction (QEC), the detrimental
effects of bit flips and phase flips on data qubits can quickly render
a quantum computer useless. Many different quantum error correction
codes are known \cite{Got97,Ste96,LMPZ96,BK98,Kni05,Sho95}. Each error
correction code has different advantages and disadvantages. The
five-qubit code \cite{Sho95} is the smallest QEC code able to correct
an arbitrary single-qubit error. Although it is a good candidate for
quantum memory protection, it is difficult to manipulate data encoded
in the five-qubit error correction code in a fault-tolerant
manner. This is because there are no transversal multiple-qubit
logical gates, and few transversal single-qubit gates. In contrast,
the seven-qubit Steane code \cite{Ste96,LMPZ96} \emph{is} able to
perform a variety of transversal logical gates, including a
transversal CNOT gate, and several transversal single-qubit
rotations. However, the seven-qubit code is less efficient than the
five-qubit code at storing information, using two additional data
qubits to encode a single logical qubit.

There are situations in which it is desirable to change the encoding
of qubits stored by a quantum computer. For example, in may be
desirable to have information stored in long term memory encoded using
one error correction code, and the information being manipulated by
the processor encoded using a different error correction code. The
requirements for memory might be focused on using \emph{small} codes,
whereas those for the processor focused on ease of
operation. Similarly, information being transmitted down a bus might
be optimized to mitigate errors due to loss, but this might not be a
relevant requirement for a processor. It is therefore necessary to
have efficient, fault-tolerant methods to convert between different
error correction codes.

In this paper, we demonstrate how data encoded in one quantum error
correction code can be fault-tolerantly converted into
another. Specifically, we show how data encoded in a five-qubit QEC
code can be converted, efficiently and fault-tolerantly, into a
seven-qubit Steane code.

Conversion between error correction codes is achieved by progressing
through a series of codes, each of which is a valid error correction
code in its own right. Each of these codes is only slightly different
from the last. Each code can not only correct any single qubit error,
but also any extra errors which might be introduced by the conversion
operations. If error correction is applied at each step of the
conversion, the encoded information remains protected.

There are many different ways to convert between codes. The method we
present in this paper was found through computational search to use
the fewest number of two-qubit control-sign ($CZ$) operations to
convert between codes. We obtained the optimal conversion between the
two codes by performing a breadth-first search from codes locally
equivalent to the five-qubit code and proceeded to codes locally
equivalent to the seven-qubit code. This allowed us to find a path of
minimal length between the two codes.

This paper is organized as follows: Section \ref{sec:method} describes
the method used to convert between codes. The QEC
codes in this section are enumerated in the appendix. Section
\ref{sec:loc} considers the initial and final codes, and demonstrates
that they are the five- and seven-qubit codes respectively. Discussion
of the computational search is given in Section \ref{sec:optimal} and
the resources required are discussed in Section \ref{sec:res}. Finally, a
conclusion is given in Section \ref{sec:conclusion}.

\section{Method} \label{sec:method}

This section describes the conversion from the five-qubit code to the
seven-qubit code. We assume that the information is initially encoded
in the five-qubit code, and we would like this information to be
encoded in the seven-qubit code. For the purpose of this paper, we
allow a total of ten data qubits for storing the data --- three
ancilla qubits above the seven needed to encode data in the seven-qubit code, and five more than is required for the five-qubit
code. Initially, each of these five ancilla qubits is assumed to be
initialized in the $\ket{+}$ state. In addition to these ten data
qubits, additional qubits are required to fault-tolerantly determine
the syndrome measurements. These qubits will be discussed in more
detail in Section \ref{sec:res}.

Throughout this paper, we use $X$, $Y$ and $Z$ to refer to the Pauli
matrices $\sigma_X$, $\sigma_Y$ and $\sigma_Z$, as well as $I$ to
refer to the identity. In writing out the stabilizers for a
multi-qubit code or state, we omit the tensor product (e.g. $X \otimes
X$ is written as $XX$). The Hadamard gate is defined as $H =
(X+Z)/\sqrt{2}$. We will also generically refer to the generators of a
given stabilizer group for a given code as `the stabilizers' of the
code.

In order to convert between codes, we assume that we are able to
perform two basic operations. These are:

\begin{description}
\item[Application of the $CZ$ Gate] We allow the application of the
  controlled-sign gate between any two qubits. This well known gate is
  a two-qubit entangling operation and applies a $\pi$ phase to the
  $\ket{11}$ state, while not affecting any of the other states. That
  is
\begin{equation}
 CZ = \left[
\begin{array}{cccc}
 1 & 0 & 0 & 0 \\
 0 & 1 & 0 & 0 \\
 0 & 0 & 1 & 0 \\
 0 & 0 & 0 & -1
\end{array}
\right].
\end{equation}

\item[Single-qubit operations]

Single-qubit operations (in particular, the Hadamard gate, and $X$ and
$Z$ rotations) can be applied in parallel to the qubits.

\end{description}

\begin{figure}
\includegraphics[width=8cm]{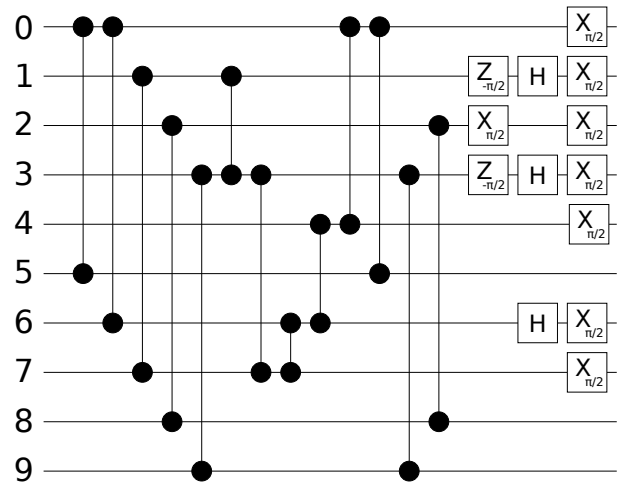}
\caption{Circuit diagram of conversion between error correction
  codes. Solid circles with a line connecting them represent
  control-sign ($CZ$) gates. Single-qubit rotations are labelled by the
  axis of rotation and a subscripted angle of rotation. Hadamard gates
  are labelled by `$H$'.} \label{fig:circuit}
\end{figure}

In order to convert a five-qubit code to a seven-qubit code, several
different operations are performed, giving rise to slightly different
QEC codes at each step. At each step, we will explicitly consider both
the stabilizers of the code, and the logical operations. A circuit of
the operations which are applied to convert from the five-qubit code
to the seven-qubit code is shown in Figure \ref{fig:circuit}. After
each operation, appropriate stabilizers are fault-tolerantly measured
and, if required, corrections may be applied.

If a $CZ$ gate is applied, it introduces the possibility of errors on
two different qubits. One situation where this can happen is when a
single qubit error occurs during the operation of the $CZ$ gate. For
our purposes, we assume the worst possible scenario: that two-qubit
gate can cause any combination of one or two Pauli errors on the
qubits affected by the gate. In order to be fault-tolerant, therefore,
we ensure the subsequent code is able to identify and correct both
one-qubit errors (on any qubit), and two-qubit errors on both ends of
an applied $CZ$ gate.

The application of a stabilizer does not affect the state at
all. Therefore if two errors differ by only a stabilizer of a state,
their effect on the state is equivalent. Similarly, they may be
corrected by applying the same operation to the quantum state. We do
not need to be able to distinguish such errors, and refer to the two
errors as equivalent `modulo the stabilizers'. Errors which are not
equivalent modulo the stabilizers, we will call `distinct'.

It is possible to verify that each of these codes is a valid error
correction code. This is achieved by enumerating all errors and
verifying that they produce a unique syndrome, or are equivalent
errors modulo the stabilizers.  The sets of stabilizers for each code
(enumerated in the appendix) were checked (by computer) and verified
to give a unique syndrome for all distinct sets of errors. Both single
qubit errors and two-qubit errors were considered.

The final QEC code is locally equivalent to the
seven-qubit Steane code. Information which was initially encoded in
the five-qubit error correction code has been fault-tolerantly
transferred to be protected by the seven-qubit Steane code.

\section{Local equivalence of the codes} \label{sec:loc}

In this section, we show that the initial and final codes are locally
equivalent to the five- and seven-qubit codes respectively. The
initial and final codes are not expressed in the usual sets of
stabilizers found in the literature for five- and seven-qubit
codes. Here we explicitly give the local operations which convert
between the forms used in the paper.

The initial code is locally equivalent to the five-qubit code. Its
stabilizers are similar to the traditional five-qubit code
\cite{Got97}:

\begin{center}
\begin{tabular}{c}
\hline
Stabilizers \\
\hline
$
\begin{array}{cccccc}
 Y & Y & Z & I & Z\\ 
 Z & Y & Y & Z & I\\ 
 I & Z & Y & Y & Z\\ 
 Z & I & Z & Y & Y\\ 

\end{array}
$ \\
\hline
\end{tabular}
\end{center}

The logical operators for the five-qubit QEC
code are:
\begin{center}
\begin{tabular}{c}
\hline
Logical Operators \\
\hline
$
\begin{array}{ccccccccc}
X_L =&  X & X & X & X & X\\ 
Z_L =&  Z & Z & Z & Z & Z\\ 
\end{array}
$ \\
\hline
\end{tabular}
\end{center}

The only difference is a change from $X$ to $Y$ and qubit reordering,
which has no effect other than to change which syndrome is associated
with which error. For the purposes of this paper, we use this
trivially changed version of the error correction code.

In the final stage of the code conversion, a set of single-qubit
rotations is applied. There is no reason why these operations cannot
be applied in parallel. This operation is fault-tolerant.

The final code is slightly more difficult to verify. After the
completion of the $CZ$ gates and the first three local operations (not
including the Hadamard gate), the code is as follows:

\begin{center}
\begin{tabular}{c}
\hline
Stabilizers \\
\hline
$
\begin{array}{ccccccccccc}
 X & I & X & Z & I & I & Z & I & I & I\\ 
 I & Z & X & I & X & I & Z & I & I & I\\ 
 Z & X & Z & I & I & I & I & Z & I & I\\ 
 I & I & Z & X & Z & I & I & Z & I & I\\ 
 I & I & I & I & I & X & I & I & I & I\\ 
 Z & I & I & I & Z & I & X & Z & I & I\\ 
 I & Z & I & Z & I & I & Z & X & I & I\\ 
 I & I & I & I & I & I & I & I & X & I\\ 
 I & I & I & I & I & I & I & I & I & X\\ 

\end{array}
$ \\
\hline
\end{tabular}
\end{center}

The logical operators for the QEC code are:
\begin{center}
\begin{tabular}{c}
\hline
Logical Operators \\
\hline
$
\begin{array}{ccccccccccccc}
X_L =& Y & X & X & X & Y & I & I & I & I & I\\ 
Z_L =& Z & Z & Y & Z & Z & I & I & I & I & I\\ 
\end{array}
$ \\
\hline
\end{tabular}
\end{center}

After the application of the Hadamard gates to qubits 1, 3 and 6, and
the removal of unentangled qubits 9, 8 and 5, the stabilizers of the
code become:

\begin{center}
\begin{tabular}{c}
\hline
Stabilizers \\
\hline
$
\begin{array}{ccccccccccc}
 X & I & X & X & I & X & I \\ 
 I & X & X & I & X & X & I \\ 
 I & X & I & X & I & X & X \\ 
 Z & Z & Z & I & I & I & Z \\ 
 I & I & Z & Z & Z & I & Z \\ 
 Z & I & I & I & Z & Z & Z \\ 
\end{array}
$ \\
\hline
\end{tabular}
\end{center}

with the following logical operators,
\begin{center}
\begin{tabular}{c}
\hline
Logical Operators \\
\hline
$
\begin{array}{ccccccccccc}
X_L =&  X & X & X & X & X & X & X \\ 
Z_L =&  Y & Y & Y & Y & Y & Y & Y 
\end{array}
$ \\
\hline
\end{tabular}
\end{center}

The code stabilizers are made up of only combinations of $Z$ and $X$
terms alone. Rearranging the order of these terms, and the qubits
\begin{eqnarray*}
 0 & \leftarrow & 0 \\
 1 & \leftarrow & 1 \\
 2 & \leftarrow & 6 \\
 3 & \leftarrow & 2 \\
 4 & \leftarrow & 4 \\
 5 & \leftarrow & 3 \\
 6 & \leftarrow & 5 \\
\end{eqnarray*}
and rearranging the stabilizers gives:
\begin{center}
\begin{tabular}{c}
\hline
Stabilizers \\
\hline
$
\begin{array}{ccccccccccc}
 X & X & X & X & I & I & I \\ 
 X & X & I & I & X & X & I \\ 
 X & I & X & I & X & I & X \\ 
 Z & Z & Z & Z & I & I & I \\ 
 Z & Z & I & I & Z & Z & I \\ 
 Z & I & Z & I & Z & I & Z \\ 
\end{array}
$ \\
\hline
\end{tabular}
\end{center}
with the following logical operators,
\begin{center}
\begin{tabular}{c}
\hline
Logical Operators \\
\hline
$
\begin{array}{ccccccccccc}
X_L =& X & X & X & X & X & X & X \\ 
Z_L =& Y & Y & Y & Y & Y & Y & Y  
\end{array}
$ \\
\hline
\end{tabular}
\end{center}

Although the logical $Z_L$ operation (at this point) is not the
standard $Z^{\otimes 7}$, this can easily remedied by applying an
$X_{\pi/2}$ gate to each of the seven qubits. After this operation,
the stabilizers remain the same, but the logical operators become:

\begin{center}
\begin{tabular}{c}
\hline
Logical Operators \\
\hline
$
\begin{array}{ccccccccccc} 
X_L =&  X & X & X & X & X & X & X\\
Z_L =&  Z & Z & Z & Z & Z & Z & Z
\end{array}
$ \\
\hline
\end{tabular}
\end{center}

\section{Optimal solution} \label{sec:optimal}

Figure \ref{fig:circuit} is an optimal solution --- that is, it uses
the fewest number of two-qubit operations. In order to find this
solution, we used a computer program, which searched through the space
of possibilities as follows:

Starting at both the five and the seven-qubit codes, the program
exhaustively enumerates all locally equivalent graph states
corresponding to the encoded logical operators of each code
\cite{HEB04, CSS+09}. The search then proceeded by applying all
possible $CZ$ gates to each of these codes. Each resulting code is
then checked to see if it is a valid quantum error correction code,
and only accepted if it able to correct any relevant error at that
point. The algorithm proceeds as a breadth-first search until the
first collision is found between codes originating from both the five-
and seven-qubit codes. When such a collision is found, we have found a
shortest length path from the five to the seven-qubit codes.

In order to keep the number of error correction codes searched to a
manageable size, it was important to identify codes which are
isomorphic to each other. Many codes are isomorphic simply because of
a rearrangement of qubits. Although, in general, identifying all
isomorphic codes is difficult, some codes can be quickly be recognised
as isomorphic to each other. These were cached, and not searched
twice.

Similarly, each path through a series of codes is isomorphic to other
QEC codes by multiplication of single-qubit rotations. These
operations do not change the distance of the error correction code,
and so only one such representative path was considered.

Using this method we were able to find paths of shortest length
between the five-qubit and seven-qubit QEC codes.

The solution we found was restricted to use exactly ten data
qubits. For differing numbers of available data qubits, different
solutions might be possible. For example, a solution with eleven, or
nine qubits might be obtainable.

Often we will also want to convert the seven-qubit code into the
five-qubit code. In general, it is not possible to simply reverse the
order of a set of steps converting one code to another. This is
because of the extra two-qubit errors which need to be accounted for
when two-qubit operations are applied. In the reverse direction it is
a \emph{different} set of stabilizers, and a different error
correction code which has to account for these extra errors. For this
particular set of codes, it is possible to run all the steps in
reverse direction as well: the error correction code (in the reverse
direction) can also account for the relevant two-qubit errors. Since
our solution was found only checking one of the two directions (away
from the five-qubit code for the first half of the operations or away
from the seven-qubit codes for the second), the solution we found is
the optimum solution for the more stringent requirement that the
conversion be reversible.

\section{Resources Required} \label{sec:res}

We now consider the resources required to fault-tolerantly transfer a
five-qubit code to a seven-qubit code. In the procedure above, we
explicitly used codes of up to ten data qubits to encode the
information. In addition to these, ancilla qubits are required to make
syndrome measurements. In the scheme we have presented, stabilizers of
up to weight 6 need to be measured (although there are stabilizers of
weight eight listed in the appendix, combinations of these are
equivalent to stabilizers of weight six). This then requires seven
additional qubits, six of which are prepared in a cat state, in order
to perform fault-tolerant measurement of the syndrome \cite{Sho95}. In
total then, at any one time we require a maximum of 17 qubits
including ten data qubits and seven ancilla qubits for measurement of
the syndrome.

Including only operations modifying the error correction codes (as
opposed to counting adding or removing qubits, syndrome measurement
and error correction), the number of operations required to convert
between codes is $15$, made up of 14 $CZ$ gates, and one application
of single-qubit gates in parallel ($3$ Hadamard gates, $8$ $X$
rotations and $2$ $Z$ rotations).

\section{Conclusion} \label{sec:conclusion}

We have shown a new, optimal, method to fault-tolerantly convert
between the five-qubit and seven-qubit Steane QEC codes. The
conversion works by changing through a series of quantum error
correction codes, each slightly distinct from the last. Each of these
codes can correct both single-qubit errors, as well as any two-qubit
errors which might be introduced by two-qubit operations. We have
shown that each step was valid by explicitly calculating the
stabilizers of the code, and verifying that every syndrome produced
(modulo the stabilizers of the code) is unique.

This research was conducted by the Australian Research Council Centre
of Excellence for Quantum Computation and Communication Technology
(project number CE110001027), with support from the US National
Security Agency and the US Army Research Office under contract number
W911NF-08-1-0527.

\bibliographystyle{plain}
\bibliography{bibliography}

\section*{Appendix: Explicit codes} \label{sec:appendix}

In this appendix we explicitly list the stabilizers and the logical
operators for each of the codes.

The following operations were performed:

\begin{enumerate}
\item Initial state

The following are the stabilizers for the QEC code:

\begin{center}
\begin{tabular}{c}
\hline
Stabilizers \\
\hline
$
\begin{array}{cccccc}
 Y & Y & Z & I & Z\\ 
 Z & Y & Y & Z & I\\ 
 I & Z & Y & Y & Z\\ 
 Z & I & Z & Y & Y\\ 

\end{array}
$ \\
\hline
\end{tabular}
\end{center}

The logical operators for the QEC code are:
\begin{center}
\begin{tabular}{c}
\hline
Logical Operators \\
\hline
$
\begin{array}{cccccccc}
X_L =&  X & X & X & X & X\\ 
Z_L =&  Z & Z & Z & Z & Z\\ 

\end{array}
$ \\
\hline
\end{tabular}
\end{center}

\item Apply a $CZ$ operation between qubits 0 and 5.

The following are the stabilizers for the QEC code:

\begin{center}
\begin{tabular}{c}
\hline
Stabilizers \\
\hline
$
\begin{array}{ccccccccccc}
 Y & Y & Z & I & Z & Z & I & I & I & I\\ 
 Z & Y & Y & Z & I & I & I & I & I & I\\ 
 I & Z & Y & Y & Z & I & I & I & I & I\\ 
 Z & I & Z & Y & Y & I & I & I & I & I\\ 
 Z & I & I & I & I & X & I & I & I & I\\ 
 I & I & I & I & I & I & X & I & I & I\\ 
 I & I & I & I & I & I & I & X & I & I\\ 
 I & I & I & I & I & I & I & I & X & I\\ 
 I & I & I & I & I & I & I & I & I & X\\ 

\end{array}
$ \\
\hline
\end{tabular}
\end{center}

The logical operators for the QEC code are:
\begin{center}
\begin{tabular}{c}
\hline
Logical Operators \\
\hline
$
\begin{array}{cccccccccccccc}
X_L =&  X & X & X & X & X & Z & I & I & I & I\\ 
Z_L =&  Z & Z & Z & Z & Z & I & I & I & I & I\\ 

\end{array}
$ \\
\hline
\end{tabular}
\end{center}
\item Apply a $CZ$ operation between qubits 0 and 6.

The following are the stabilizers for the QEC code:

\begin{center}
\begin{tabular}{c}
\hline
Stabilizers \\
\hline
$
\begin{array}{ccccccccccc}
 Y & Y & Z & I & Z & Z & Z & I & I & I\\ 
 Z & Y & Y & Z & I & I & I & I & I & I\\ 
 I & Z & Y & Y & Z & I & I & I & I & I\\ 
 Z & I & Z & Y & Y & I & I & I & I & I\\ 
 Z & I & I & I & I & X & I & I & I & I\\ 
 Z & I & I & I & I & I & X & I & I & I\\ 
 I & I & I & I & I & I & I & X & I & I\\ 
 I & I & I & I & I & I & I & I & X & I\\ 
 I & I & I & I & I & I & I & I & I & X\\ 

\end{array}
$ \\
\hline
\end{tabular}
\end{center}

The logical operators for the QEC code are:
\begin{center}
\begin{tabular}{c}
\hline
Logical Operators \\
\hline
$
\begin{array}{cccccccccccccc}
X_L =&  X & X & X & X & X & Z & Z & I & I & I\\ 
Z_L =&  Z & Z & Z & Z & Z & I & I & I & I & I\\ 

\end{array}
$ \\
\hline
\end{tabular}
\end{center}
\item Apply a $CZ$ operation between qubits 1 and 7.

The following are the stabilizers for the QEC code:

\begin{center}
\begin{tabular}{c}
\hline
Stabilizers \\
\hline
$
\begin{array}{ccccccccccc}
 Y & Y & Z & I & Z & Z & Z & Z & I & I\\ 
 Z & Y & Y & Z & I & I & I & Z & I & I\\ 
 I & Z & Y & Y & Z & I & I & I & I & I\\ 
 Z & I & Z & Y & Y & I & I & I & I & I\\ 
 Z & I & I & I & I & X & I & I & I & I\\ 
 Z & I & I & I & I & I & X & I & I & I\\ 
 I & Z & I & I & I & I & I & X & I & I\\ 
 I & I & I & I & I & I & I & I & X & I\\ 
 I & I & I & I & I & I & I & I & I & X\\ 

\end{array}
$ \\
\hline
\end{tabular}
\end{center}

The logical operators for the QEC code are:
\begin{center}
\begin{tabular}{c}
\hline
Logical Operators \\
\hline
$
\begin{array}{cccccccccccccc}
X_L =&  X & X & X & X & X & Z & Z & Z & I & I\\ 
Z_L =&  Z & Z & Z & Z & Z & I & I & I & I & I\\ 

\end{array}
$ \\
\hline
\end{tabular}
\end{center}
\item Apply a $CZ$ operation between qubits 2 and 8.

The following are the stabilizers for the QEC code:

\begin{center}
\begin{tabular}{c}
\hline
Stabilizers \\
\hline
$
\begin{array}{ccccccccccc}
 Y & Y & Z & I & Z & Z & Z & Z & I & I\\ 
 Z & Y & Y & Z & I & I & I & Z & Z & I\\ 
 I & Z & Y & Y & Z & I & I & I & Z & I\\ 
 Z & I & Z & Y & Y & I & I & I & I & I\\ 
 Z & I & I & I & I & X & I & I & I & I\\ 
 Z & I & I & I & I & I & X & I & I & I\\ 
 I & Z & I & I & I & I & I & X & I & I\\ 
 I & I & Z & I & I & I & I & I & X & I\\ 
 I & I & I & I & I & I & I & I & I & X\\ 

\end{array}
$ \\
\hline
\end{tabular}
\end{center}

The logical operators for the QEC code are:
\begin{center}
\begin{tabular}{c}
\hline
Logical Operators \\
\hline
$
\begin{array}{cccccccccccccc}
X_L =&  X & X & X & X & X & Z & Z & Z & Z & I\\ 
Z_L =&  Z & Z & Z & Z & Z & I & I & I & I & I\\ 

\end{array}
$ \\
\hline
\end{tabular}
\end{center}
\item Apply a $CZ$ operation between qubits 3 and 9.

The following are the stabilizers for the QEC code:

\begin{center}
\begin{tabular}{c}
\hline
Stabilizers \\
\hline
$
\begin{array}{ccccccccccc}
 Y & Y & Z & I & Z & Z & Z & Z & I & I\\ 
 Z & Y & Y & Z & I & I & I & Z & Z & I\\ 
 I & Z & Y & Y & Z & I & I & I & Z & Z\\ 
 Z & I & Z & Y & Y & I & I & I & I & Z\\ 
 Z & I & I & I & I & X & I & I & I & I\\ 
 Z & I & I & I & I & I & X & I & I & I\\ 
 I & Z & I & I & I & I & I & X & I & I\\ 
 I & I & Z & I & I & I & I & I & X & I\\ 
 I & I & I & Z & I & I & I & I & I & X\\ 

\end{array}
$ \\
\hline
\end{tabular}
\end{center}

The logical operators for the QEC code are:
\begin{center}
\begin{tabular}{c}
\hline
Logical Operators \\
\hline
$
\begin{array}{cccccccccccccc}
X_L =&  X & X & X & X & X & Z & Z & Z & Z & Z\\ 
Z_L =&  Z & Z & Z & Z & Z & I & I & I & I & I\\ 
\end{array}
$ \\
\hline
\end{tabular}
\end{center}
\item Apply a $CZ$ operation between qubits 1 and 3.

The following are the stabilizers for the QEC code:

\begin{center}
\begin{tabular}{c}
\hline
Stabilizers \\
\hline
$
\begin{array}{ccccccccccc}
 Y & Y & Z & Z & Z & Z & Z & Z & I & I\\ 
 Z & Y & Y & I & I & I & I & Z & Z & I\\ 
 I & I & Y & Y & Z & I & I & I & Z & Z\\ 
 Z & Z & Z & Y & Y & I & I & I & I & Z\\ 
 Z & I & I & I & I & X & I & I & I & I\\ 
 Z & I & I & I & I & I & X & I & I & I\\ 
 I & Z & I & I & I & I & I & X & I & I\\ 
 I & I & Z & I & I & I & I & I & X & I\\ 
 I & I & I & Z & I & I & I & I & I & X\\ 

\end{array}
$ \\
\hline
\end{tabular}
\end{center}

The logical operators for the QEC code are:
\begin{center}
\begin{tabular}{c}
\hline
Logical Operators \\
\hline
$
\begin{array}{cccccccccccccc}
X_L =&  X & Y & X & Y & X & Z & Z & Z & Z & Z\\ 
Z_L =&  Z & Z & Z & Z & Z & I & I & I & I & I\\ 

\end{array}
$ \\
\hline
\end{tabular}
\end{center}
\item Apply a $CZ$ operation between qubits 7 and 3.

The following are the stabilizers for the QEC code:

\begin{center}
\begin{tabular}{c}
\hline
Stabilizers \\
\hline
$
\begin{array}{ccccccccccc}
 Y & Y & Z & Z & Z & Z & Z & Z & I & I\\ 
 Z & Y & Y & I & I & I & I & Z & Z & I\\ 
 I & I & Y & Y & Z & I & I & Z & Z & Z\\ 
 Z & Z & Z & Y & Y & I & I & Z & I & Z\\ 
 Z & I & I & I & I & X & I & I & I & I\\ 
 Z & I & I & I & I & I & X & I & I & I\\ 
 I & Z & I & Z & I & I & I & X & I & I\\ 
 I & I & Z & I & I & I & I & I & X & I\\ 
 I & I & I & Z & I & I & I & I & I & X\\ 

\end{array}
$ \\
\hline
\end{tabular}
\end{center}

The logical operators for the QEC code are:
\begin{center}
\begin{tabular}{c}
\hline
Logical Operators \\
\hline
$
\begin{array}{cccccccccccccc}
X_L =&  X & Y & X & Y & X & Z & Z & I & Z & Z\\ 
Z_L =&  Z & Z & Z & Z & Z & I & I & I & I & I\\ 

\end{array}
$ \\
\hline
\end{tabular}
\end{center}
\item Apply a $CZ$ operation between qubits 7 and 6.

The following are the stabilizers for the QEC code:

\begin{center}
\begin{tabular}{c}
\hline
Stabilizers \\
\hline
$
\begin{array}{ccccccccccc}
 Y & Y & Z & Z & Z & Z & Z & Z & I & I\\ 
 Z & Y & Y & I & I & I & I & Z & Z & I\\ 
 I & I & Y & Y & Z & I & I & Z & Z & Z\\ 
 Z & Z & Z & Y & Y & I & I & Z & I & Z\\ 
 Z & I & I & I & I & X & I & I & I & I\\ 
 Z & I & I & I & I & I & X & Z & I & I\\ 
 I & Z & I & Z & I & I & Z & X & I & I\\ 
 I & I & Z & I & I & I & I & I & X & I\\ 
 I & I & I & Z & I & I & I & I & I & X\\ 

\end{array}
$ \\
\hline
\end{tabular}
\end{center}

The logical operators for the QEC code are:
\begin{center}
\begin{tabular}{c}
\hline
Logical Operators \\
\hline
$
\begin{array}{cccccccccccccc}
X_L =&  X & Y & X & Y & X & Z & Z & I & Z & Z\\ 
Z_L =&  Z & Z & Z & Z & Z & I & I & I & I & I\\ 
\end{array}
$ \\
\hline
\end{tabular}
\end{center}
\item Apply a $CZ$ operation between qubits 6 and 4.

The following are the stabilizers for the QEC code:

\begin{center}
\begin{tabular}{c}
\hline
Stabilizers \\
\hline
$
\begin{array}{ccccccccccc}
 Y & Y & Z & Z & Z & Z & Z & Z & I & I\\ 
 Z & Y & Y & I & I & I & I & Z & Z & I\\ 
 I & I & Y & Y & Z & I & I & Z & Z & Z\\ 
 Z & Z & Z & Y & Y & I & Z & Z & I & Z\\ 
 Z & I & I & I & I & X & I & I & I & I\\ 
 Z & I & I & I & Z & I & X & Z & I & I\\ 
 I & Z & I & Z & I & I & Z & X & I & I\\ 
 I & I & Z & I & I & I & I & I & X & I\\ 
 I & I & I & Z & I & I & I & I & I & X\\ 

\end{array}
$ \\
\hline
\end{tabular}
\end{center}

The logical operators for the QEC code are:
\begin{center}
\begin{tabular}{c}
\hline
Logical Operators \\
\hline
$
\begin{array}{cccccccccccccc}
X_L =&  X & Y & X & Y & X & Z & I & I & Z & Z\\ 
Z_L =&  Z & Z & Z & Z & Z & I & I & I & I & I
\end{array}
$ \\
\hline
\end{tabular}
\end{center}
\item Apply a $CZ$ gate between qubits 4 and 0.

The following are the stabilizers for the QEC code:

\begin{center}
\begin{tabular}{c}
\hline
Stabilizers \\
\hline
$
\begin{array}{ccccccccccc}
 Y & Y & Z & Z & I & Z & Z & Z & I & I\\ 
 Z & Y & Y & I & I & I & I & Z & Z & I\\ 
 I & I & Y & Y & Z & I & I & Z & Z & Z\\ 
 I & Z & Z & Y & Y & I & Z & Z & I & Z\\ 
 Z & I & I & I & I & X & I & I & I & I\\ 
 Z & I & I & I & Z & I & X & Z & I & I\\ 
 I & Z & I & Z & I & I & Z & X & I & I\\ 
 I & I & Z & I & I & I & I & I & X & I\\ 
 I & I & I & Z & I & I & I & I & I & X\\ 

\end{array}
$ \\
\hline
\end{tabular}
\end{center}

The logical operators for the QEC code are:
\begin{center}
\begin{tabular}{c}
\hline
Logical Operators \\
\hline
$
\begin{array}{cccccccccccccc}
X_L =&  Y & Y & X & Y & Y & Z & I & I & Z & Z\\ 
Z_L =&  Z & Z & Z & Z & Z & I & I & I & I & I\\ 

\end{array}
$ \\
\hline
\end{tabular}
\end{center}
\item Apply a $CZ$ gate between qubits 5 and 0.

The following are the stabilizers for the QEC code:

\begin{center}
\begin{tabular}{c}
\hline
Stabilizers \\
\hline
$
\begin{array}{ccccccccccc}
 Y & Y & Z & Z & I & I & Z & Z & I & I\\ 
 Z & Y & Y & I & I & I & I & Z & Z & I\\ 
 I & I & Y & Y & Z & I & I & Z & Z & Z\\ 
 I & Z & Z & Y & Y & I & Z & Z & I & Z\\ 
 I & I & I & I & I & X & I & I & I & I\\ 
 Z & I & I & I & Z & I & X & Z & I & I\\ 
 I & Z & I & Z & I & I & Z & X & I & I\\ 
 I & I & Z & I & I & I & I & I & X & I\\ 
 I & I & I & Z & I & I & I & I & I & X\\ 

\end{array}
$ \\
\hline
\end{tabular}
\end{center}

The logical operators for the QEC code are:
\begin{center}
\begin{tabular}{c}
\hline
Logical Operators \\
\hline
$
\begin{array}{cccccccccccccc}
X_L =&  Y & Y & X & Y & Y & I & I & I & Z & Z\\ 
Z_L =&  Z & Z & Z & Z & Z & I & I & I & I & I\\ 
\end{array}
$ \\
\hline
\end{tabular}
\end{center}
\item Apply a $CZ$ gate between qubits 9 and 3.

The following are the stabilizers for the QEC code:

\begin{center}
\begin{tabular}{c}
\hline
Stabilizers \\
\hline
$
\begin{array}{ccccccccccc}
 Y & Y & Z & Z & I & I & Z & Z & I & I\\ 
 Z & Y & Y & I & I & I & I & Z & Z & I\\ 
 I & I & Y & Y & Z & I & I & Z & Z & I\\ 
 I & Z & Z & Y & Y & I & Z & Z & I & I\\ 
 I & I & I & I & I & X & I & I & I & I\\ 
 Z & I & I & I & Z & I & X & Z & I & I\\ 
 I & Z & I & Z & I & I & Z & X & I & I\\ 
 I & I & Z & I & I & I & I & I & X & I\\ 
 I & I & I & I & I & I & I & I & I & X\\ 

\end{array}
$ \\
\hline
\end{tabular}
\end{center}

The logical operators for the QEC code are:
\begin{center}
\begin{tabular}{c}
\hline
Logical Operators \\
\hline
$
\begin{array}{cccccccccccccc}
X_L =&  Y & Y & X & Y & Y & I & I & I & Z & I\\ 
Z_L =&  Z & Z & Z & Z & Z & I & I & I & I & I\\ 

\end{array}
$ \\
\hline
\end{tabular}
\end{center}
\item Apply a $CZ$ gate between qubits 8 and 2. 

The following are the stabilizers for the QEC code:

\begin{center}
\begin{tabular}{c}
\hline
Stabilizers \\
\hline
$
\begin{array}{ccccccccccc}
 Y & Y & Z & Z & I & I & Z & Z & I & I\\ 
 Z & Y & Y & I & I & I & I & Z & I & I\\ 
 I & I & Y & Y & Z & I & I & Z & I & I\\ 
 I & Z & Z & Y & Y & I & Z & Z & I & I\\ 
 I & I & I & I & I & X & I & I & I & I\\ 
 Z & I & I & I & Z & I & X & Z & I & I\\ 
 I & Z & I & Z & I & I & Z & X & I & I\\ 
 I & I & I & I & I & I & I & I & X & I\\ 
 I & I & I & I & I & I & I & I & I & X\\ 

\end{array}
$ \\
\hline
\end{tabular}
\end{center}

The logical operators for the QEC code are:
\begin{center}
\begin{tabular}{c}
\hline
Logical Operators \\
\hline
$
\begin{array}{cccccccccccccc}
X_L =&  Y & Y & X & Y & Y & I & I & I & I & I\\ 
Z_L =&  Z & Z & Z & Z & Z & I & I & I & I & I\\ 

\end{array}
$ \\
\hline
\end{tabular}
\end{center}
\item Applied single-qubit operation $X_\frac{\pi}{2}$ to qubit 2 and $Z_{-\frac{\pi}{2}}$ to qubits 1 and 3.

The following are the stabilizers for the QEC code:

\begin{center}
\begin{tabular}{c}
\hline
Stabilizers \\
\hline
$
\begin{array}{ccccccccccc}
 X & I & X & Z & I & I & Z & I & I & I\\ 
 I & Z & X & I & X & I & Z & I & I & I\\ 
 Z & X & Z & I & I & I & I & Z & I & I\\ 
 I & I & Z & X & Z & I & I & Z & I & I\\ 
 I & I & I & I & I & X & I & I & I & I\\ 
 Z & I & I & I & Z & I & X & Z & I & I\\ 
 I & Z & I & Z & I & I & Z & X & I & I\\ 
 I & I & I & I & I & I & I & I & X & I\\ 
 I & I & I & I & I & I & I & I & I & X\\ 
\end{array}
$ \\
\hline
\end{tabular}
\end{center}

The logical operators for the QEC code are:
\begin{center}
\begin{tabular}{c}
\hline
Logical Operators \\
\hline
$
\begin{array}{cccccccccccccc}
X_L = & Y & X & X & X & Y & I & I & I & I & I\\ 
Z_L = & Z & Z & Y & Z & Z & I & I & I & I & I\\ 
\end{array}
$ \\
\hline
\end{tabular}
\end{center}

\end{enumerate}

\end{document}